\documentclass[a4paper,11pt]{article}
\begin{document}

\title{\bf Semantics of Programming Languages:\\
A Tool-Oriented Approach\thanks{This research was supported in part by
the Telematica Instituut under the {\em Domain-Specific Languages} project.}}

\author{{Jan Heering}\\
        {\em\footnotesize CWI}\\
        {\em\footnotesize P.O. Box~94079, 1090~GB Amsterdam, The Netherlands}\\
        {\tt\footnotesize Jan.Heering@cwi.nl}
        \and
        {Paul Klint}\\
        {\em\footnotesize CWI and University of Amsterdam}\\
        {\em\footnotesize P.O. Box~94079, 1090~GB Amsterdam, The Netherlands}\\
        {\tt\footnotesize Paul.Klint@cwi.nl}
       }      

\date{}

\maketitle
\begin{abstract}
By paying more attention to semantics-based tool generation,
programming language semantics can significantly increase its
impact. Ultimately, this may lead to ``Language Design Assistants''
incorporating substantial amounts of semantic knowledge.
\end{abstract}

\section{The Role of Programming Language Semantics} \label{sec:ROLE}

Programming language semantics has lost touch with large groups of
potential users \cite{Schmidt96}. Among the reasons for this
unfortunate state of affairs, one stands out. Semantic results are
rarely incorporated in practical systems that would help language
designers to implement and test a language under development, or
assist programmers in answering their questions about the meaning of
some language feature not properly documented in the language's
reference manual. Nevertheless, such systems are potentially more
effective in bringing semantics-based formalisms and techniques to the
places they are needed than their dissemination in publications,
courses, or even exemplary (but little-used) programming languages.

\begin{figure}
\rule{17cm}{.5mm}
\begin{picture}(400,220)(-240,-200)
% last tuple = position

\thicklines

\put(0,-60)
  {\begin{picture}(200,200)
      \put(0,0){\oval(60,30)}
      \put(0,0){\makebox(0,0){\bf Semantics}}
      \put(0,20){\vector(0,1){50}}
  \end{picture}
  }

\put(-60,-150)
  {\begin{picture}(200,200)
      \put(0,0){\oval(60,30)}
      \put(0,0){\makebox(0,0){\bf Languages}}
      \put(-27,-20){\vector(-1,-1){15}}
  \end{picture}
  }

\put(60,-150)
  {\begin{picture}(200,200)
      \put(0,0){\oval(60,30)}
      \put(0,0){\makebox(0,0){\bf Tools}}
      \put(27,-20){\vector(1,-1){15}}
  \end{picture}
  }

\end{picture}
\normalsize
\rule{17cm}{.5mm}
\caption{Semantics, languages, and tools are drifting steadily further apart.} \label{fig:CURRENT}

\rule{17cm}{.5mm}
\begin{picture}(400,220)(-240,-200)
% last tuple = position

\thicklines

\put(0,-60)
  {\begin{picture}(200,200)
      \put(0,0){\oval(60,30)}
      \put(0,0){\makebox(0,0){\bf Semantics}}
      \put(0,-20){\vector(0,-1){15}}
  \end{picture}
  }

\put(-60,-150)
  {\begin{picture}(200,200)
      \put(0,0){\oval(60,30)}
      \put(0,0){\makebox(0,0){\bf Languages}}
      \put(27,15){\vector(1,1){11}}
  \end{picture}
  }

\put(60,-150)
  {\begin{picture}(200,200)
      \put(0,0){\oval(60,30)}
      \put(0,0){\makebox(0,0){\bf Tools}}
      \put(-27,15){\vector(-1,1){11}}
  \end{picture}
  }

\put(0,-110)
  {\begin{picture}(200,200)
      \put(0,0){\oval(60,30)}
      \put(0,0){\makebox(0,0){\bf TG/LDAs}}
  \end{picture}
  }

\end{picture}
\normalsize
\rule{17cm}{.5mm}
\caption{In the tool-oriented approach, semantics, languages, and tools
are kept together by Tool Generation ({\bf TG}) and, ultimately, Language Design
Assistants ({\bf LDAs}).} \label{fig:VISION}
\end{figure}

The current situation in which semantics, languages, and tools are
drifting steadily further apart is shown in Figure~\ref{fig:CURRENT}.
The tool-oriented approach to semantics aims at
making semantics definitions more useful and productive by generating
as many language-based tools from them as possible.  This will, we
expect, reverse the current trend as shown in Figure~\ref{fig:VISION}.
The goal is to produce semantically well-founded languages and tools.
Ultimately, we envision the emergence of ``Language Design
Assistants'' incorporating substantial amounts of semantic knowledge.

\begin{table}[t]
\begin{center}
\begin{tabular}{|p{5cm}|p{5cm}|}
\hline
Semantics          & Definition in terms of      \\ 
\hline \hline

Axiomatic \cite{AptOlderog97} & Pre- and postconditions \\

Attribute grammars \cite{DJL88} & Attribute propagation rules \\

Denotational       \cite{Schmidt86}  & Lambda-expressions          \\ 

Algebraic          \cite{BHK89} & Equations/rewrite rules     \\ 

Structural operational \cite{Plotkin81}/ Natural \cite{Kahn87} & Inference rules \\ 

Action \cite{Mosses92}  & Action expressions          \\ 

Abstract state machines \cite{Gurevich95} & Transition rules       \\ 

Coalgebraic  \cite{JR97} &  Behavioral specification rules \\ 

Program algebra \cite{BergstraLoots98} & Equations \\

\hline
\end{tabular}
\caption{Current approaches to programming language semantics.} \label{table:METHODS}
\end{center}
\end{table}

Table~\ref{table:METHODS} lists the semantics definition methods we
are aware of.  Examples of their use can be found in
\cite{SK95}. Petri nets, process algebras, and other methods that do
not specifically address the semantics of programming languages, are
not included.  Dating back to the sixties, attribute grammars and
denotational semantics are among the oldest methods, while abstract
state machines (formerly called evolving algebras), coalgebra
semantics, and program algebra are the latest additions to the field.
Ironically, while attribute grammars are popular with tool builders,
semanticists do not consider them a particularly interesting
definition method.  Since we will only discuss the various methods in
general terms without going into technical details, the reader need
not be familiar with them. In any case, the differences between them,
while often hard to decipher because the field is highly fragmented
and appropriate ``dictionaries'' are lacking, do not affect our main
argument.

\begin{table}
\begin{center}
\begin{tabular}{|p{4cm}|p{4.5cm}|p{6cm}|}
\hline
Semantics & System & Developed at    \\ 
\hline \hline

Attribute grammars & Synthesizer Generator \cite{RT89} & Cornell University \\

Denotational       & PSG \cite{BahlkeSnelting86} & Technical University of Darmstadt \\
              
Algebraic          & ASF+SDF Meta-En\-vi\-ron\-ment \cite{DHK96} & CWI and University of Amsterdam \\

Structural operational/ Natural  & Centaur \cite{BorrasEtAl89} & INRIA Sophia-Anti\-po\-lis \\

Action             & ASD \cite{vanDeursen94} & CWI and University of Aarhus\\

Abstract state machi\-nes & Gem-Mex \cite{AnlauffEtAl97} & University of L'Aquila \\

(Operational) & Software Refinery \cite{MarkosianEtAl94} & Reasoning Systems, Palo Alto \\

\hline
\end{tabular}
\caption{Some representative language development systems.} \label{table:SYSTEMS}
\end{center}
\end{table}

Table~\ref{table:SYSTEMS} lists a representative language development
system (if any) for the semantics definition methods of
Table~\ref{table:METHODS}. The last entry, Software Refinery, which
has its origins in knowledge-based software environments research at
Kestrel Institute, does not fit any of the current semantics
paradigms. The pioneering Semanol system \cite{AndersonEtAl76} is, to
the best of our knowledge, no longer in use and is not included.  The
systems listed have widely different capabilities and are in widely
different stages of development. Before discussing their
characteristics and applications in Section~\ref{sec:SYSTEMS}, we
first explain the general ideas underlying the tool-oriented approach
to programming language semantics.  These were shaped by our
experiences with the ASF+SDF Meta-Environment
(Table~\ref{table:SYSTEMS}) over the past ten years.  Finally, we
discuss Language Design Assistants in Section~\ref{sec:LDA}.

\section{A Tool-Oriented Approach to Semantics} \label{sec:APPROACH}

The tool-oriented approach to semantics aims at making semantics
definitions more useful and productive by generating as many
language-based tools from them as possible. This affects many
aspects of the way programming language semantics is practiced and
upsets some of its dogmas.

\begin{table}[t]
\begin{center}
\begin{tabular}{|c|}
 \hline
        Scanner/Parser \\
        Prettyprinter \\
        Syntax-directed editor \\
        Typechecker \\
        (Abstract) interpreter(s) \\
        Dataflow analyzer \\
        Call graph extractor \\
        Partial evaluator \\
        Optimizer \\
        Program slicer  \\
        Origin tracker \\ 
        Debugger \\
        Code generator \\
        Compiler  \\
        Profiler \\
        Test case generator \\
        Test coverage analyzer \\
        Regression test tool \\
        Complexity analyzer (metrics) \\
        Documentation generator \\
        Cluster analysis tool \\
        Systematic program modification tool \\
\hline
\end{tabular}
\caption{Tools that might be derived from a language definition.} \label{table:TOOLS}
\end{center}
\end{table}

Table~\ref{table:TOOLS} lists some of the tools that might be
generated.  In principle, the language definition has to be augmented
with suitable tool-specific information for each tool to be generated,
and this may require tool-specific language extensions to the core
semantics definition formalism.  In practice, this is not always
necessary since semantics definitions tend to contain a good deal of
implicit information that may be extracted and used for tool
generation.  

The first entry of Table~\ref{table:TOOLS}, scanner and parser
generation, is standard technology.  Lex and Yacc are well-known
examples of stand-alone generators for this purpose.  Their input
formalisms are close to regular expressions and BNF, the
\textit{de facto} standard formalisms for regular and context-free
grammars, respectively.  Unfortunately, for most of the other tools in
Table~\ref{table:TOOLS} there are no such standard formalisms.

The key features of the tool-oriented approach are:
\begin{itemize}

\item Language definitions are primarily tool generator input.
      They do not have to provide any kind of theoretical
      ``explanation'' of the constructs of the language in question
      nor do they have to become part of a language reference manual.

\item An interpreter that can act, among other things, as an ``oracle''
      to programmers needing help will be among the first tools to be
      generated.

\item Writing (large) language definitions loses its esoteric
      character and becomes similar to any other kind of programming.
      Semantics formalisms tend to do best on small examples, but lose
      much of their power as the language definitions being written
      grow.  In the tool-oriented approach, semantics formalisms have
      to be modular and separate generation (the analogue of separate
      compilation) has to be supported. Libraries of language
      constructs become important.

\item The tool-oriented approach may require addition of tool-specific
      features to the core formalism.  This leads to an open-ended
      rather than a ``pure'' style of semantics description.
  
\item The scope of the tool-oriented approach includes, for instance,
 \begin{itemize} 

 \item Domain-specific and little languages
       \cite{vanDeursenKlint98,Salus98}. Many of the tools in
       Table~\ref{table:TOOLS} are as useful for DSLs as they are
       for programming languages.

 \item Software maintenance and renovation tools \cite{vanDeursenEtAl99}.
       Some of these are included at the end of Table~\ref{table:TOOLS}.

 \item Compiler toolkits such as CoSy \cite{Ace99}, Cocktail
       \cite{GroschEmmelmann90}, OCS \cite{JusticeBudd93}, SUIF
       \cite{WilsonEtAl94}, and PIM \cite{BDFH97,FieldEtAl98}.

 \end{itemize} 

\end{itemize}

\section{Existing Language Development Systems} \label{sec:SYSTEMS}

\begin{table}[t]
\begin{center}
\begin{tabular}{|p{3cm}|p{10cm}|p{2cm}|}
\hline

System                         & Generated tools 
                               & Semantic engine \\ \hline \hline
 
Synthesizer Generator          & scanner/parser (LALR), prettyprinter,
                                 syntax-directed editor, incremental typechecker,
                                 incremental translator, $\ldots$
                               & incremental attribute evaluator
\\ \hline

PSG                            & scanner/parser, syntax-directed editor,
                                 incremental typechecker (even for incomplete
                                 program fragments), interpreter                 
                               & functional language interpreter
\\ \hline
              
ASF+SDF Meta-En\-vi\-ron\-ment & scanner/parser (generalized LR), 
                                 prettyprinter, syntax-directed editor,
                                 typechecker, interpreter,
                                 origin tracker, translator,
                                 renovation tools, $\ldots$
                               & conditional rewrite rule engine
\\ \hline

Centaur                        & scanner/parser (LALR),
                                 prettyprinter, syntax-directed editor,
                                 typechecker, interpreter,
                                 origin tracker, translator, $\ldots$ 
                               & inference rule engine   
\\ \hline

ASD                            & scanner/parser, syntax-directed editor, 
                                 checker, interpreter
                               & conditional rewrite rule engine
\\ \hline

Gem-Mex                        & scanner/parser, typechecker, interpreter,
                                 debugger
                               & transition rule engine
\\ \hline

Software Refinery              & scanner/parser (LALR), prettyprinter,
                                 syntax-directed editor, object-oriented
                                 parse tree repository (including dataflow
                                 relations), 
                                 Y2K/Euro tools, program slicer, $\ldots$
                               & tree manipulation engine 
\\ \hline

\end{tabular}
\caption{Tool generation capabilities of representative language development systems.} \label{table:CAPABILITIES}
\end{center}
\end{table}

Table~\ref{table:CAPABILITIES} summarizes the tool generation
capabilities of the representative language development systems listed
in Table~\ref{table:SYSTEMS}.  All of them can generate lexical
scanners, parsers, and prettyprinters, many of them can produce
syntax-directed editors, typecheckers, and interpreters, and a few can
produce various kinds of software renovation tools.  To this end, they
support one or more specification formalisms, but these differ in
generality and application domain.  

For instance, the Synthesizer Generator supports attribute grammars
with incremental attribute evaluation, which is particularly suitable
for typechecking, static analysis and translation, but less suitable
for dynamic semantics.  The ASF+SDF Meta-Environment supports
conditional rewrite rules rather than attribute grammars, and these
can be used for defining dynamic semantics as well.  Software Refinery
comes with a full-blown functional language in which a wide range of
computations on programs can be expressed.  Other systems provide more
specialized specification formalisms. PSG, for instance, uses context
relations to describe incremental typechecking (even for incomplete
program fragments) and denotational definitions for dynamic semantics.
Gem-Mex supports a semi-visual formalism optimized for the definition
of programming language semantics and tool generation. It can generate
a typechecker, an interpreter, and a debugger.

Table~\ref{table:CAPABILITIES} is far from complete.  Some other
language development systems are SIS \cite{Mosses79}, PSP
\cite{Paulson82}, GAG \cite{Kastens84}, SPS \cite{Wand84},
MESS \cite{Lee89}, Actress \cite{BrownEtAl92},
Pregmatic \cite{vandenBrand92}, LDL \cite{HarmEtAl97},
and Eli \cite{KastensEtAl98}.
Many of the tools listed in Table~\ref{table:TOOLS} are not generated
by any current system.  Ample opportunities for tool generation still
exist in areas like optimization, dynamic program analysis, testing,
and maintenance.

\section{Toward Language Design Assistants} \label{sec:LDA}

The logical next step beyond semantics-based tool generation would
lead to a situation similar to that of computer algebra. Large parts
of mathematics are being incorporated in computer algebra
systems. Conversely, computer algebra itself has become a fruitful
mathematical activity, yielding new results of general mathematical
interest.  In the case of semantics, we see opportunities for
``Language Design Assistants'' incorporating a substantial amount of
both formal and informal semantic knowledge. The latter is found, for
instance, in language design rationales and discussion documents
produced by standardization bodies.  Development of such assistants
will not only push semantics even further toward practical
application, but also give rise to new theoretical questions.

The Language Design Assistants we have in mind would support the human
language designer by providing design choices and performing
consistency checks during the design process.  Operational knowledge
about typical issues like typing rules, scope rules, and execution
models should be incorporated in them.  Major research questions arise
here regarding the acquisition, representation, organization, and
abstraction level of the required knowledge. For instance, should it
be organized according to any of the currently known paradigms of
object-oriented, functional, or logic programming? Or should a higher
level of abstraction be found from which these and other, new,
paradigms can be derived? How can constraints on the composition of
certain features be expressed and checked?  Another key question is
how to construct a collection of ``language feature
components'' that are sufficiently general to be reusable across a
wide range of languages.

Similar considerations apply to tool development. By incorporating
knowledge about tool generation in the Language Design Assistant we
can envision a Tool Generation Assistant that helps in constructing
tools in a more advanced way than the tool generation we had in mind
in the previous sections.

\begin{table}[t]
\rule{17cm}{.5mm}
\begin{center}
\begin{minipage}{15cm}
\begin{description}

\item[$\bigcirc$] What is the type of the expression controlling the selection
      of one of the two branches.

\item[$\bigcirc$] How is the controlling expression evaluated (short circuit
      vs. full evaluation)?

\item[$\bigcirc$] Is the controlling expression evaluated concurrently with other
      program parts (with speculative execution of the conditional as
      a special case)?

\item[$\bigcirc$] Can the controlling expression have side-effects?

\item[$\bigcirc$] Can the controlling expression cause exceptions?

\item[$\bigcirc$] Are jumps from outside into the branches allowed?

\item[$\bigcirc$] Is the selected branch evaluated concurrently with other
      program parts?

\item[$\bigcirc$] Can the evaluation of the selected branch cause side-effects?

\item[$\bigcirc$] Can the evaluation of the selected branch cause exceptions?

\item[$\bigcirc$] Does the evaluation of the conditional construct yield a value?

\end{description}
\end{minipage}
\end{center}
\rule{17cm}{.5mm}
\caption{Some of the possible parameters of a generic conditional construct.} \label{table:IF}
\end{table}

To make this perspective somewhat more tangible, consider the
relatively simple case of an if-then-else-like conditional construct
that has to be modelled as a language feature component.
Table~\ref{table:IF} gives an impression of the wide range of issues
that has to be addressed before such a generic conditional construct
can be specialized into a concrete if-then-else-statement or
conditional expression in a specific language.  It is a research
question to design an abstract framework in which these and similar
questions can be expressed and answered.

Another major question is how to organize the specialization process
from language feature component to concrete language construct. The
main alternatives are parameterization and transformation
\cite{Laemmel99}.  Using parameterization, specialization of the
component in question amounts to instantiating its parameters.  Since
parameters have to be identified beforehand and instantiation is
usually a rather simple mechanism, the adaptability/reusability of a
parameterized component is limited.  Using transformations, on the
other hand, a language feature component is designed without explicit
parameters.  Specialization is achieved by applying appropriate
transformation rules to it to obtain the desired specific case.
Clearly, this approach is more flexible since any part of the language
feature component can be modified by the transformation rules and can
thus effectively act as a parameter.  The relation between this
approach of meta-level transformation and parameterized modules is
largely unexplored.

Although we are not aware of research on Language Design Assistants
from the broad perspective sketched here, there is some work pointing
in the same general direction:
\begin{itemize}

\item The Language Designer's Workbench sketched as future work
      in \cite{Pleban84,Lee89} has some of the same goals.

\item Action semantics \cite{Mosses92} also emphasizes libraries
      of reusable language constructs.

\item Plans (no longer pursued) for the Language Development Laboratory
      \cite{HarmEtAl97} included a library of reusable language
      constructs, a knowledge base containing knowledge of languages
      and their compilers/interpreters, and a tool for language
      design.

\item The ``design and implementation by selection'' of languages
      described in \cite{PfahlerKastens97,KastensPfahler98} is a case
      study in high-level interactive composition of predefined
      language constructs.

\end{itemize}

\paragraph{Acknowledgements} We would like to thank Jan Bergstra,
Mark van den Brand, Arie van Deursen, Ralf L\"{a}mmel, and Jan Rutten
for useful comments on earlier versions.

% \small 
\bibliographystyle{abbrv}

\def\bibname{References}
  \addcontentsline{toc}{chapter}{\protect\numberline{}\bibname}
  \newcommand{\sortunder}[1]{}

\end{document}